\documentclass[preprint,12pt]{elsarticle}

\usepackage{graphicx}
\usepackage{subfigure,amsmath,amssymb}
\usepackage{mathptmx}      
\usepackage{url}

\newtheorem{finding}{Finding}

\newcommand{\M}{\mathcal{M}}

\begin{document}

\title{Phenomenology of retained refractoriness: \\
On semi-memristive discrete media}
\author{Andrew Adamatzky$^1$ and Leon O. Chua$^2$} 
\address{$^1$ University of the West of England,  Bristol, UK \\
\url{andrew.adamatzky@uwe.ac.uk}\\
            $^2$ University of California at Berkeley, Berkeley, USA \\
\url{chua@eecs.berkeley.edu}
}

\begin{abstract}
We study two-dimensional cellular automata, each cell takes three states: resting, excited and refractory. 
A resting cell excites if number of excited neighbours lies in a certain interval (excitation interval). An excited cell become refractory independently on states of its neighbours. A refractory cell returns to a resting state only if the number of excited neighbours belong to recovery interval. The model is an excitable cellular automaton abstraction of  a spatially extended semi-memristive medium where a cell's resting state symbolises low-resistance and refractory state high-resistance. The medium is \emph{semi}-memristive because only transition from high- to low-resistance is controlled by density of local excitation. We present phenomenological classification of the automata behaviour for all possible excitation intervals and recovery intervals. We describe eleven classes of  cellular automata with retained refractoriness based on criteria of space-filling ratio,  morphological and generative diversity,
and types of travelling localisations.

\vspace{0.5cm}
\noindent
\emph{Keywords:}  cellular automaton, excitable medium, space-time dynamics

\end{abstract}

\maketitle

\section{Introduction}

The memristor (a passive resistor with memory) is a device whose resistance changes depending on the polarity and magnitude of a voltage applied to the device's terminals and the duration of this voltage's application. Its existence was theoretically postulated by Leon Chua in 1971 based on symmetry in integral variations of OhmÕs laws~\cite{chua:1971,chua:1976,chua:1980}. The memristor is characterised by a non-linear relationship between the charge and the flux; this relationship can be generalised to any two-terminal device in which resistance depends on the internal state of the system~\cite{chua:1976}. The memristor cannot be implemented using the three other passive circuit elements --- resistor, capacitor and inductor --Ð therefore the memristor is an atomic element of electronic circuitry~\cite{chua:1971,chua:1976,chua:1980}. Using memristors one can achieve circuit functionalities that it is not possible to establish with resistors, capacitors and inductors, therefore the memristor is of great pragmatic usefulness.  The first experimental prototypes of memristors are reported in~\cite{williams:2008,erokhin:2008,yang:2008}.  Potential unique applications of memristors are in spintronic devices, ultra-dense information storage, neuromorphic circuits, and programmable electronics~\cite{strukov:2008}.

Despite explosive growth of results in  memristor studies there is still a few (if any!) findings on phenomenology of spatially extended non-linear media with hundreds of thousands of locally connected memristors. We attempt to fill the gap and develop a minimalistic model of a discrete  memristive medium.  The only --- so far --- approaches to develop cellular automata model of a memristive medium are Itoh-Chua memristor cellular automata, where cellular automaton lattice is actually designed of memristors~\cite{itoh:2009}, and Adamatzky-Chua model of memrisitive cellular automata based on structurally-dynamic cellular automata~\cite{adamatzky_memristive_excitable}. Both models imitate memristive properties of links, connections between cells of automata arrays but not the cells themselves. In this paper  we explore the scenario when link between cells are always 'conductive' but cells themselves can take non-conductive (refractory) states.

We define semi-memristive automata, or excitable cellular automata with retained refractoriness, in Sect.~\ref{definition}.  Section~\ref{classes} presents grouping of cell-state transition functions into eleven classes. Hierarchies of classes based on their morphological diversity, 
space-filling ratio and expressiveness are constructed in Sect.~\ref{hierarchies}.
In Sect.~\ref{localisations} we overview a 'zoo' of travelling localizations observed in simulated cellular automata and 
exemplify interactions between the localizations. Overview of the results is provided in Sect.~\ref{discussion}.

\section{Definitions and methods}
\label{definition}

\subsection{Semi-memristive automaton}

A cellular automaton $\mathcal{A}$ is an orthogonal array of uniform finite-state machines, or cells. Each cell takes the finite number of states and updates its states in discrete time depending on states of its closest neighbours. All cells update their states simultaneously by the same rule. We consider eight-cell neighbourhood and three cell-states: resting $\circ$, excited $+$, and refractory $-$. Let $u(x) = \{  y: |x-y|_{L\infty}=1\}$ be a neighbourhood of cell $x$. Let $\sigma^t_x = \sum_{y \in u(x)} \chi(y^t, +)$ be the sum of excited neighbours of cell $x$ at time step $t$, where  $\chi(y^t, +)=1$ if $y^t=+$ and $\chi(y^t, +)=0$ otherwise.   
Let $\Theta=[\theta_1, \theta_2]$, $1 \leq \theta_1 \leq \theta_2  \leq 8$, be an excitation interval, and
$\Phi=[\phi_1, \phi_2]$, $1 \leq \phi_1 \leq \phi_2  \leq 8$, be a recovery interval. 
A cell $x$ updates its state by the following rule:
$$
x^{t+1}=
\begin{cases}
+, \text{ if } x^t=\circ \text{ and } \sigma^t_x \in \Theta \\
-, \text{ if } x^t =+ \\
\circ, \text{ if } (x^t=\circ  \text{ and } \sigma^t_x \not \in \Theta) \text{ or } (x^t=-  \text{ and } \sigma^t_x \in \Phi) 
\end{cases}
$$
A resting cell becomes excited if the number of its excited neighbours lies in the interval $\Theta$. An excited cell takes refractory state unconditionally. A refractory cell returns to its resting state if number of its excited neighbours lie in the interval $\Phi$. 

In computer simulations we consider $\theta_1 = 1$ or $2$ and $\theta_1 \leq \theta_2 \leq 8$; there is no point 
to consider $\theta_1 >2 $ because for such lower boundary of excitation interval most excitation patterns quickly extinct. We adopt the same boundaries of recovery interval $\Phi$:  $\phi_1 = 1, 2$ and $\phi_1 \leq \phi_2 \leq 8$. Further we refer to excitation and retention of refractoriness functions as $\M(\theta_1, \theta_2, \phi_1, \phi_2)$.

\subsection{Why semi-memristive?}

Why do we call the automaton described above a \emph{semi-memristive} automaton?  Because we assume that the refractory state  symbolises high-resistivity, or non-conductivity. A refractory cell can not be excited whatever number of excited neighbours it has, therefore it can not 'conduct' excitation.  In a fully memristive automaton both transitions from excited to refractory and from refractory to resting must be controlled by a local excitation density. In the semi-memristive automaton, the transition from excited to refractory is unconditional and happens independently of a local excitation density but the transition from a refractory state to resting state depends on a local excitation density.

\subsection{Experiments}

We experiment with cellular automaton array of $n \times n$ cells, $n=500$, and absorbing boundary conditions.
While testing automata's response to external stimulation we use point-wise and $D$-stimulation (a spatially extended stimulation in a disc). By point-wise stimulation we mean excitation of  one cell (for functions, where $\theta_1=1$) or two  neighbouring  cells (for functions, where $\theta_1=2$) of elsewhere resting cellular array. The $D$-stimulation is implemented as follows. Let $D$-disc be a set of cells which lie at distance not more than $r$ from the centre $(n/2, n/2)$ of the cellular array. When undertaking $D$-stimulation we assign excited state to a cell of $D$ with probability 0.1.
In computer experiments we used $r=100$. 

\subsection{Characteristics used for classification}

To classify morphology of configurations generated by functions  $\M(\theta_1, \theta_2, \phi_1, \phi_2)$, 
$\theta_1 = 1, 2$, $\theta_1 \leq \theta_2 \leq 8$, $\phi_1 \leq \phi_2 \leq 8$, we $D$-stimulate 
cellular automaton lattices, and subdivided them into classes by visually detected likeness of excitation and refractory patterns generated and based on the following characteristics:
\begin{itemize}
\item \emph{Space-filling ratio}. Let automaton of $n \times n$ cells being $D$-stimulated with disc radius $r$, then 
space-filling ratio is a ratio of non-resting cells in configuration recorded at time step $n-r$
(i.e. when front of propagating disturbance just reaches edge of the automaton array) to the size of array $n^2$. 
\item \emph{Morphological diversity} is the number of proportion of different neighbourhood configurations. It is evaluated using Shannon entropy and Simpson's index. Let $W=\{ \circ, +, - \}$ be a set of all possible configurations of a 9-cell neighbourhood $w(x)=u(x) \cup x$, $x \in \mathbf{L}$. Let $c$ be a configuration of automaton, we calculate number of non-resting configurations as $\eta = \sum_{ x \in \mathbf{L}} \epsilon(x)$, where $\epsilon(x)=0$ if for every resting $x$ all its neighbours are resting, and $\epsilon(x)=1$ otherwise. The Shannon entropy is calculated as $- \sum_{w \in W} (\nu(w)/\eta \cdot ln (\nu(w)/\eta))$, where $\nu(w)$ is a number of times the neighbourhood 
configuration $w$ is found in automaton configuration $c$. Simpson's index is calculated 
as $ 1- \sum_{w \in W} (\nu(w)/\eta)^2$.
\item \emph{Generative diversity} is a diversity of a configuration developed from a point-wise excitation. 
\item \emph{Expressiveness} of a function is calculated as the Shannon entropy divided by space-filling ratio. Generative expressiveness is the expressiveness calculated after point-wise stimulation.
\item \emph{Power of a class} is the number of functions in this class. 
\item \emph{Ratios of excited and refractory states} are calculated as  $n^{-2} \sum_{x \in L} \chi(x^t, +)$ and 
$n^{-2} \sum_{x \in  \mathbf{L}} \chi(x^t, -)$, where $\chi(a,b)=1$ if $a=b$, and $\chi(a,b)=0$ otherwise. 
\end{itemize}

\subsection{Classes}
\label{classes}

We separate functions $\M(\theta_1, \theta_2, \phi_1, \phi_2)$ into eleven classes, characterise morphology of configurations, generated by automata from each class, and provide details on membership of each class.

\begin{figure}[!tbp]
\centering
\subfigure[$C_1$]{\includegraphics[width=0.49\textwidth]{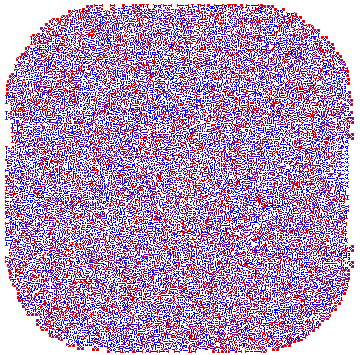}}
\subfigure[$C_2$]{\includegraphics[width=0.49\textwidth]{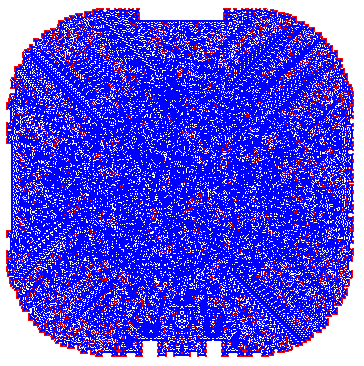}}
\subfigure[$C_3$]{\includegraphics[width=0.49\textwidth]{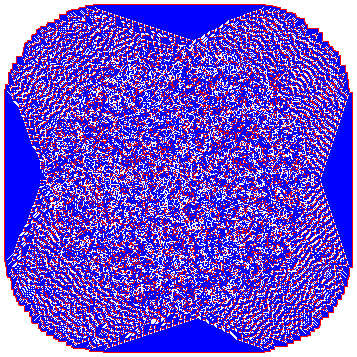}}
\subfigure[$C_4$]{\includegraphics[width=0.49\textwidth]{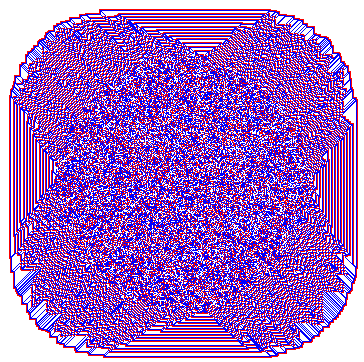}}
\caption{Exemplars of configurations generated by functions from classes $C_1$ $\ldots$ $C_4$ with $D$-stimulation. Excited cells are red (light grey), refractory cells are blue (dark grey) 
and resting cells are white.}
\label{classes_C1C4}
\end{figure}

\begin{figure}[!tbp]
\centering
\includegraphics[width=0.85\textwidth]{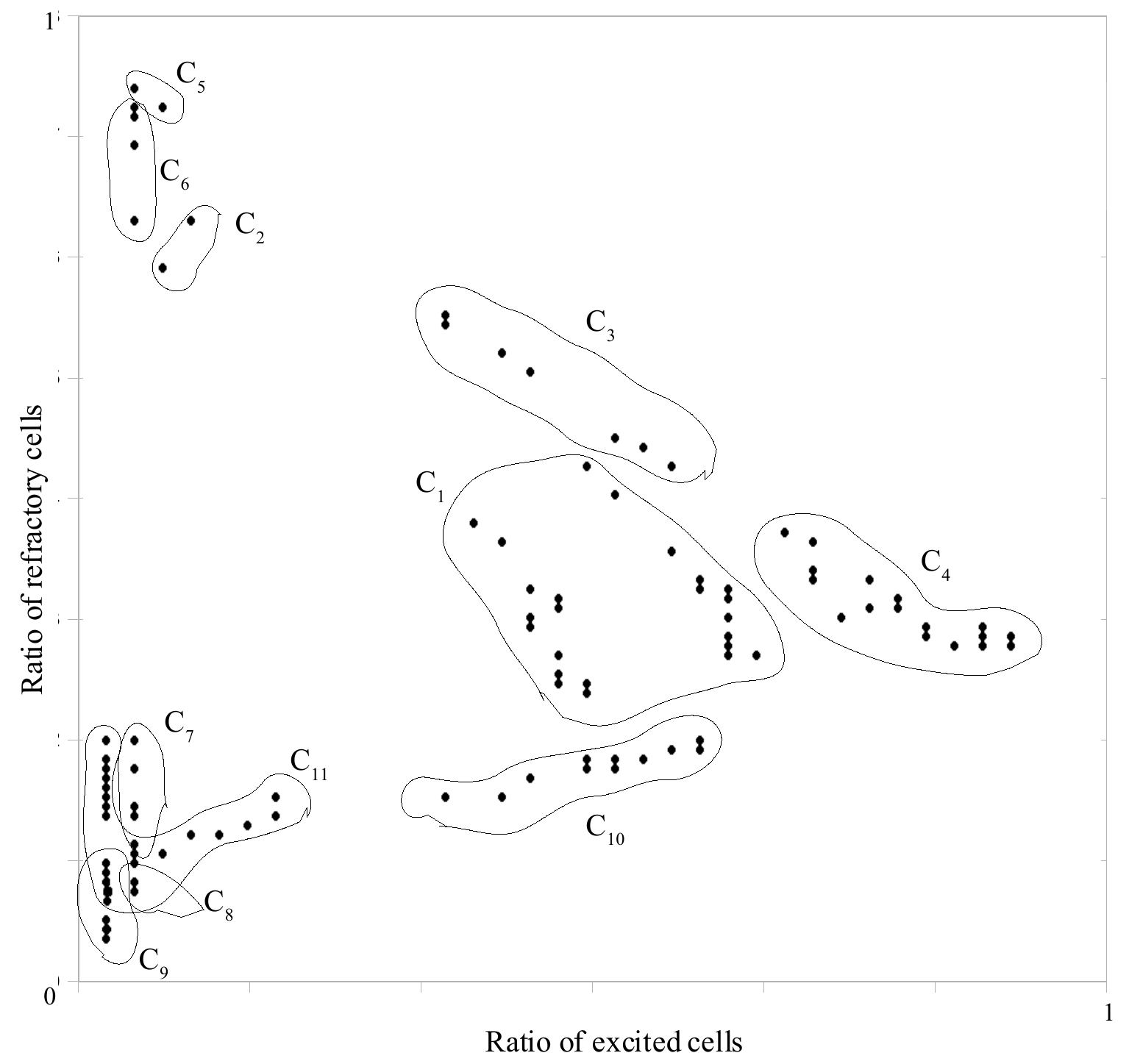}
\caption{Ratio of excited states to resting states vs ratio of refractory states to resting states in configurations developed by a cellular automat with retained refractoriness with $D$-stimulation.}
\label{ExcRef}
\end{figure}

\subsection{Class $C_1$}

$D$-stimulation gives birth to propagating quasi-chaotic pattern of excited and refractory states, 
the pattern fills the automaton array (Fig.~\ref{classes_C1C4}a). A front of the pattern is comprised of branching 
localised excitations, which periodically merge in a single excitation wave-front; the pattern is almost isotropic with slight
indentations along north-south and west-east axis. Excited and refractory states are 
exhibited equally in patterns generated by automata on class $C_1$ (Fig.~\ref{ExcRef}, $C_1$). The class $C_1$ is comprised of 27 functions:  $\M(111a)$,  $a=1,\ldots,8$; $\M(112a)$, $a=3, \ldots,8$; 
$\M(121a)$,  $a=2, \ldots 8$; $\M(122a)$,  $a=3, \ldots 8$. 

\subsection{Class $C_2$}

$D$-stimulation leads to propagating quasi-chaotic pattern with domination of refractory states. The pattern fills the automaton array (Fig.~\ref{classes_C1C4}b). The pattern's propagating front is comprised of travelling localizations, which branch periodically.  Excited localizations travelling along north-west-south-east and  north-east-south-west axis leave a distinctive trail of refractory states (Fig.~\ref{classes_C1C4}b). There is a tail of gradually extinct excitations. These spatially extended excitations collapse into localised oscillating localizations, most of which extinguish with time (Fig.~\ref{ExcRef}, $C_2$).  The class $C_2$ has three functions $\M(1122)$, $\M(1211)$, $\M(1222)$.

\subsection{Class $C_3$}

Initial random stimulation leads to  formation of a single circular excitation wave, followed by 
trains of excitation waves travelling north-west, north-east, south-west and 
south-east. The wave-trains are followed by discoidal growing quasi-random pattern of excited and refractory states (Fig.~\ref{classes_C1C4}c). North, south, east and west domains of cellular array lying between the growing patterns are 
filled entirely with refractory states. These refractory domains contribute towards slight prevalence of 
refractory states (Fig.~\ref{ExcRef}, $C_3$). The class $C_3$ has eight functions $\M(1a1b)$, $a=3, \ldots, 6$, $b=1,2$.

\subsection{Class $C_4$}

The $D$-stimulated automata response  is characterised by a single circular excitation wave, which encapsulates wave trains propagating north-west, north-east, south-west and 
south-east, and envelopes of wave-fronts travelling north, east, south and west (Fig.~\ref{classes_C1C4}d)).
The wave envelopes in $C_4$ occupying the same space domains are refractory domains in $C_3$. This is why a cluster of $C_4$ functions is positioned symmetrically to cluster of $C_3$ functions with respect to the diagonal of equal ratios of excited and refractory states in Fig.~\ref{ExcRef}.  The class $C_4$ has 48 functions $\M(1abc)$,  $a=3, \ldots, 6$, $b=1,2$, $c=3, \ldots, 8$.

\begin{figure}[!tbp]
\centering
\subfigure[$C_5$]{\includegraphics[width=0.49\textwidth]{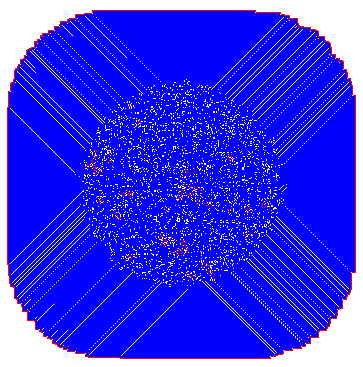}}
\subfigure[$C_6$]{\includegraphics[width=0.49\textwidth]{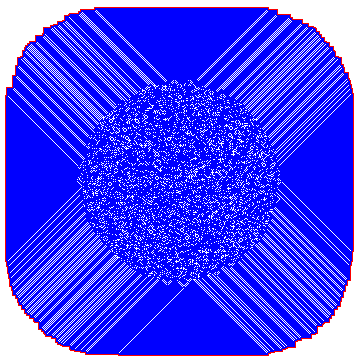}}
\subfigure[$C_7$]{\includegraphics[width=0.49\textwidth]{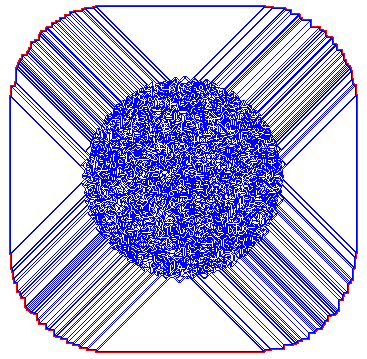}}
\subfigure[$C_8$]{\includegraphics[width=0.49\textwidth]{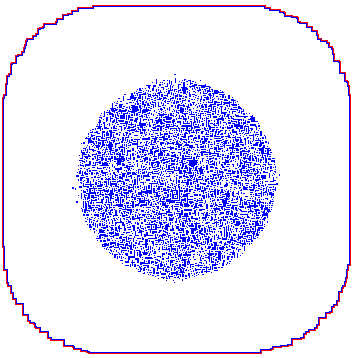}}
\caption{Exemplars of configurations generated by functions from classes $C_5$ $\ldots$ $C_8$ with $D$-stimulation. Excited cells are red (light grey), refractory cells are blue (dark grey) 
and resting cells are white.}
\label{classes_C5C8}
\end{figure}

\subsection{Classes $C_5$ and $C_6$}

In response  $D$-stimulation automata from these classes generate circular waves of excitaton  (Fig.~\ref{classes_C5C8}ab). The domain $D$ of initial perturbation is a random pattern of refractory and resting states. No excitation is observed inside $D$ in automata from class $C_6$ and usually just few localised oscillating excitations in automata from $C_5$. The circular wave-front leaves behind an almost  uniform field of refractory cells with radial traces of resting states towards north-west, north-east,  south-west and south-east directions (Fig.~\ref{classes_C5C8}ab). Functions from classes $C_5$ and $C_6$ are positioned close to class $C_2$ in  Fig.~\ref{ExcRef} due to high ratio of refractory states in configurations they generate. The class $C_5$ has four functions $\M(1a22)$, $a=3, \ldots, 6$ and the class $C_6$  has six functions  $\M(1abc)$, $a=7, 8$, $b=1, 2$, $b \leq c \leq 2$.

\subsection{Classes $C_7$ and $C_8$ }

Automata from these classes generate a single circular excitation wave while perturbed by $D$ and a domain of 
refractory and resting states inside boundaries of $D$. In automata from $C_8$ a propagating excitation wave leaves
a trail of refractory states along north-west-south-east and north-east-south-west axis, the rest of cells remain in the resting
state. Propagating wave-front leaves no traces in automata from $C_8$ (Fig.~\ref{classes_C5C8}cd). 
Class $C_7$ has 16 functions  $\M(1a1b)$, $a=7,8$, $b=3,4$;  $\M(1a2b)$, $a=7,8$, $b=3,\ldots,8$; and, class  $C_8$ has eight functions $\M(1a1b)$, $a=7,8$ and $b=5, \ldots, 8$.

\begin{figure}[!tbp]
\centering
\subfigure[$C_9$]{\includegraphics[width=0.49\textwidth]{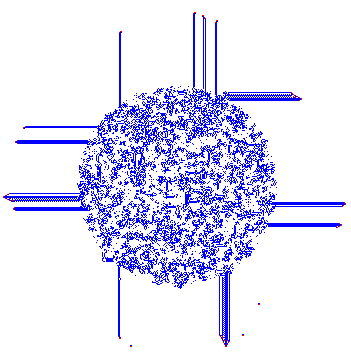}}
\subfigure[$C_{10}$]{\includegraphics[width=0.49\textwidth]{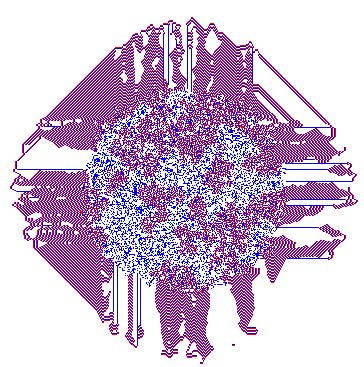}}
\subfigure[$C_{11}$]{\includegraphics[width=0.49\textwidth]{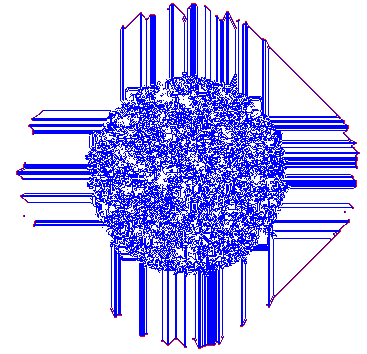}}
\caption{Exemplars of configurations generated by functions from classes $C_9$ $\ldots$ $C_{11}$ with  $D$-stimulation. Excited cells are red (light grey), refractory cells are blue (dark grey) 
and resting cells are white.}
\label{classes_C9C11}
\end{figure}

\subsection{Class $C_9$}

$D$-stimulation of an automaton from $C_9$ produces a domain of refractory and resting states 
(inside boundaries of $D$) and localizations travelling along north-south and west-east axis (Fig.~\ref{classes_C9C11}a).
The localizations leave traces of refractory states. Configurations produced by functions from $C_9$ has the lowest (amongst all other classes) ratio of excited states (Fig.~\ref{ExcRef}, $C_9$). The class $C_9$ has 17 functions  $\M(22ab)$, $a=1,2$, $a \leq b \leq 8$; $\M(2a11)$, $a=3,4$.

\subsection{Class $C_{10}$}

Phenomenology of automata's behaviour is richest amongst all classes. $D$-stimulation leads to formation of
trains of excitation wave-fragments, propagating in north-west, north-east, south-west and south-east directions;
domains of localised excitations inside boundaries of $D$, travelling localizations with and without traces of refractory 
states (Fig.~\ref{classes_C9C11}b). Typically excited cells are in slight majority in configurations generated by automata form $C_{10}$ (Fig.~\ref{ExcRef}, $C_{10}$). The class $C_{10}$ includes 61 functions  $\M(2a1b)$, $a, b =3, \ldots, 8$; $\M(2a2b)$,   $a,b=4, \ldots, 8$.

\subsection{Class $C_{11}$}

Behaviour of automata from class $C_{11}$ is somewhat similar to that of class $C_9$ but sometimes 
localizations travelling along north-south and west-east axis are 'linked' by segments from excitation 
wave-fronts (Fig.~\ref{classes_C9C11}c). The class $C_{11}$ has 27 functions $\M(2312)$; $\M(232a)$, $a=2, \ldots, 8$; $\M(2412)$;  $\M(242a)$, $a=2,3$; $\M(2abc)$, $a=5, \ldots, 8$, $b=1,2$, $b \leq c \leq 3$.

\section{Hierarchies}
\label{hierarchies}

\begin{figure}[!tbp]
\centering
\begin{footnotesize}
\begin{tabular}{c|cccc}
Class & Shannon entropy &  Simpson's index &  Space-filling ratio & Expressiveness \\ \hline
$C_1$   &	 7.01 	&  1.0 	 & 0.94  & 7.41\\
$C_2$ &	 5.06  &  0.95 	 & 0.95  &  5.35\\
$C_3$ &	 5.99 	 & 0.96 	 & 0.96  &  6.24\\
$C_4$ &	 5.55 	 & 0.98 	 & 0.96  &  5.8\\
$C_5$ &	 1.79 	 & 0.53 	 & 0.96  &  1.87\\
$C_6$ &	 2.15 	 & 0.65 	 & 0.96  &  2.23\\ 
$C_7$ &	 3.95	 & 0.96 	 & 0.42  &  9.5\\
$C_8$ &	 4.56 	 & 0.99 	 & 0.15  &  31.21\\
$C_9$ &	 3.89 	 & 0.95	 & 0.14  &  27.69\\
$C_{10}$ & 	 5.20 	 & 0.98 	 & 0.60  &  8.7\\
$C_{11}$ &   4.27 	 & 0.97 	 & 0.27  &  15.71\\
\end{tabular}
\end{footnotesize}
\caption{Average values of Shannon entropy, Simpson's index, space-filling ratio and expressiveness with $D$-stimulaton.}
\label{averages}
\end{figure}

\begin{figure}[!tbp]
\centering
\includegraphics[width=0.85\textwidth]{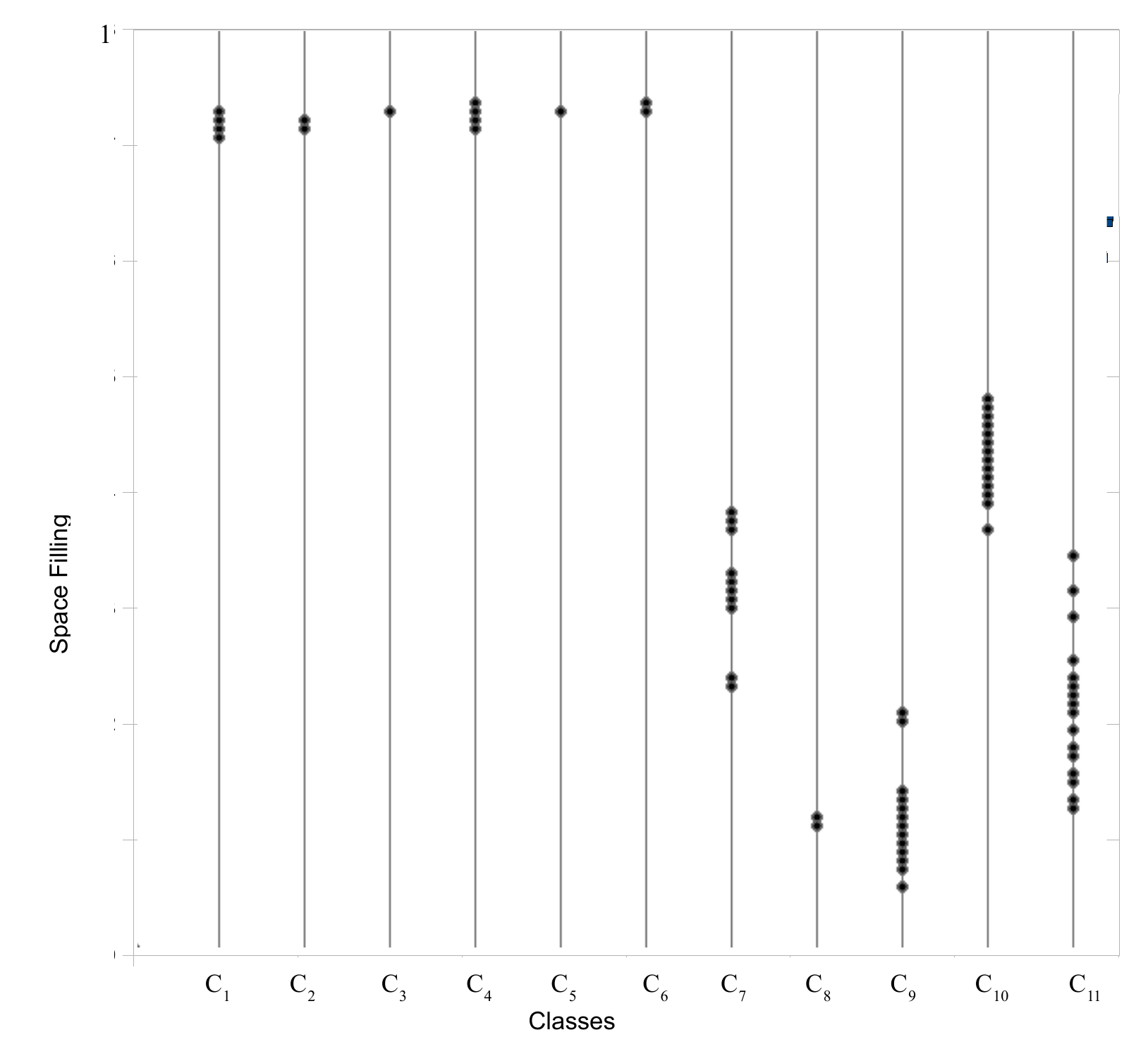}
\caption{Space-filling ratio  values for classes of retained refractoriness rules. Classes are indicated on horizontal axis.
Space-filling is shown on vertical axis. All 225 functions are mapped to show distribution of space-filling ratio inside each class. There is less discs on the chart than functions because some functions from the same class have the same space-filling ratio values.}
\label{SpaceFilling}
\end{figure}

\begin{finding}
Classes of excitable cellular automata with retained refractoriness obey the following power hierarchy: 
$$C_4 \rhd C_{10} \rhd \{ C_1, C_{11} \} \rhd C_7 \rhd C_9 \rhd \{ C_3, C_8 \} \rhd C_6 \rhd C_5 \rhd C_2 .$$
\end{finding}
This is a direct consequence of our joining functions into classes. 

\begin{finding}
Classes of excitable cellular automata with retained refractoriness obey the following hierarchy of space-filling ratio: 
$$\{ C_1, C_2, C_3, C_5, C_6 \} \rhd C_{10} \rhd C_7 \rhd C_{11} \rhd \{C_8, C_9 \}$$
\end{finding}
Space-filling ratio values averaged amongst each class are shown in Fig.~\ref{averages} and the distribution of functions inside each class in Fig.~\ref{SpaceFilling}. Classes $C_1$ to $C_6$, and $C_8$ show small variance in space-filling ratio while class $C_{11}$ shows largest variance. Roughly the classes can be split into two space-filling groups: high 
space-filling  ratio ($C_1, \ldots, C_6$) and low space-filling ratio ($C_7, \ldots, C_{11}$).

\begin{figure}[!tbp]
\centering
\subfigure[]{\includegraphics[width=0.9\textwidth]{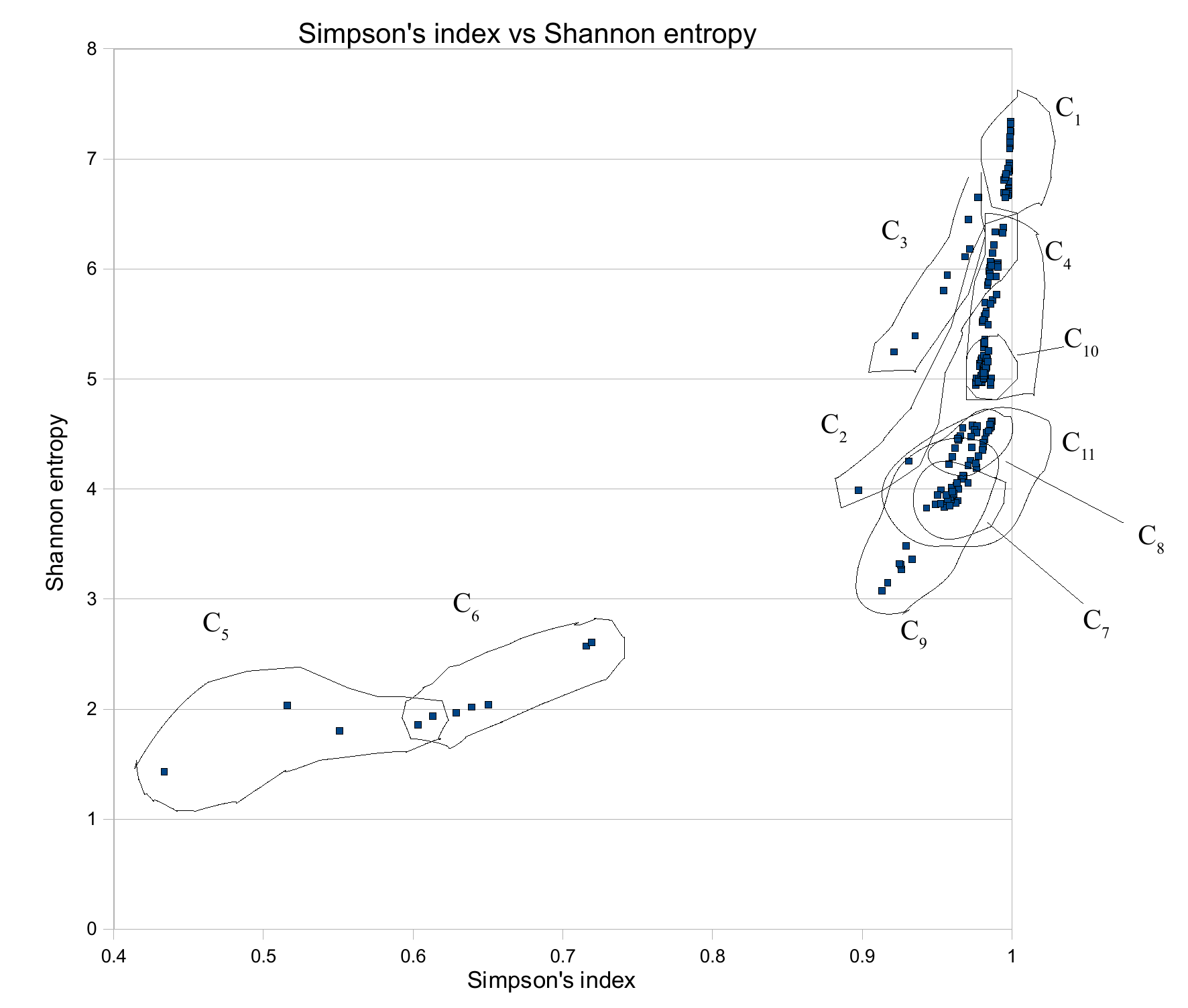}}
\subfigure[]{\includegraphics[width=0.75\textwidth]{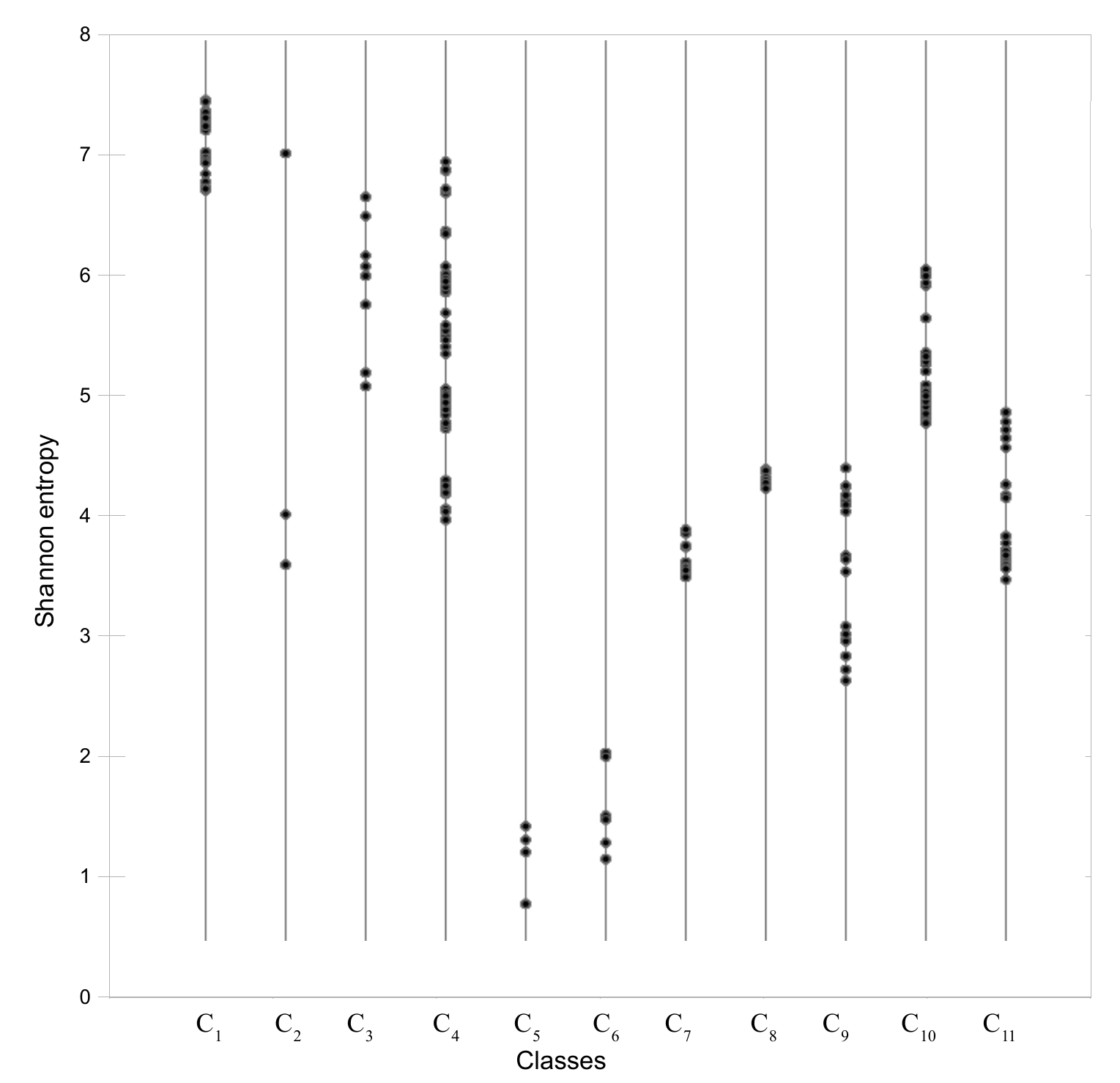}}
\caption{Distribution of classes on morphological diversity. 
(a)~Simpson's index vs Shannon entropy and (b)~Shannon entropy with $D$-stimulation.}
\label{SimpsonShannon}
\end{figure}

\begin{finding}
Classes of excitable cellular automata with retained refractoriness obey the following hierarchy of diversity: 
$$C_1 \rhd C_3 \rhd C_4 \rhd C_{10} \rhd C_2 \rhd C_8 \rhd C_{11} \rhd C_7 \rhd C_9 \rhd C_6 \rhd C_5  .$$
\end{finding}

We can draw an analogy between neighbourhood states in a given configuration of cellular automaton 
and living species in a population. Thus we can measure a diversity of a function by calculating  Simpson's index and Shannon entropy of a population of neighbourhood states in a single configuration generated by the function 
(Fig.~\ref{SimpsonShannon}). Classes $C_5$ and $C_6$ exhibit the lowest diversity because initial random excitation pattern $D$  extinguishes quickly  and only single propagating wave-front travels away from the site of original excitation.
The front leaves behind an almost homogeneous domain of refractory states. Classes $C_1$, $C_3$ and $C_4$ hold 
top ranks in the diversity hierarchy because the initial random patterns expands into a growing quasi-random configuration of 
excited and refractory states. In terms of Shannon entropy class $C_3$ shows higher degree of diversity than class $C_4$ 
(Fig.~\ref{averages}), possibly, because space-time dynamics is biased towards formation of extended wave-fronts propagating north, south, east and west, which interact with fragmentary wave-patterns travelling north-west, 
north-east, south-west and south-east. Class $C_{10}$, which demonstrates high variety of travelling localizations, and 
stays just  above the middle of the diversity hierarchy.

\begin{figure}[!tbp]
\centering
\includegraphics[width=0.85\textwidth]{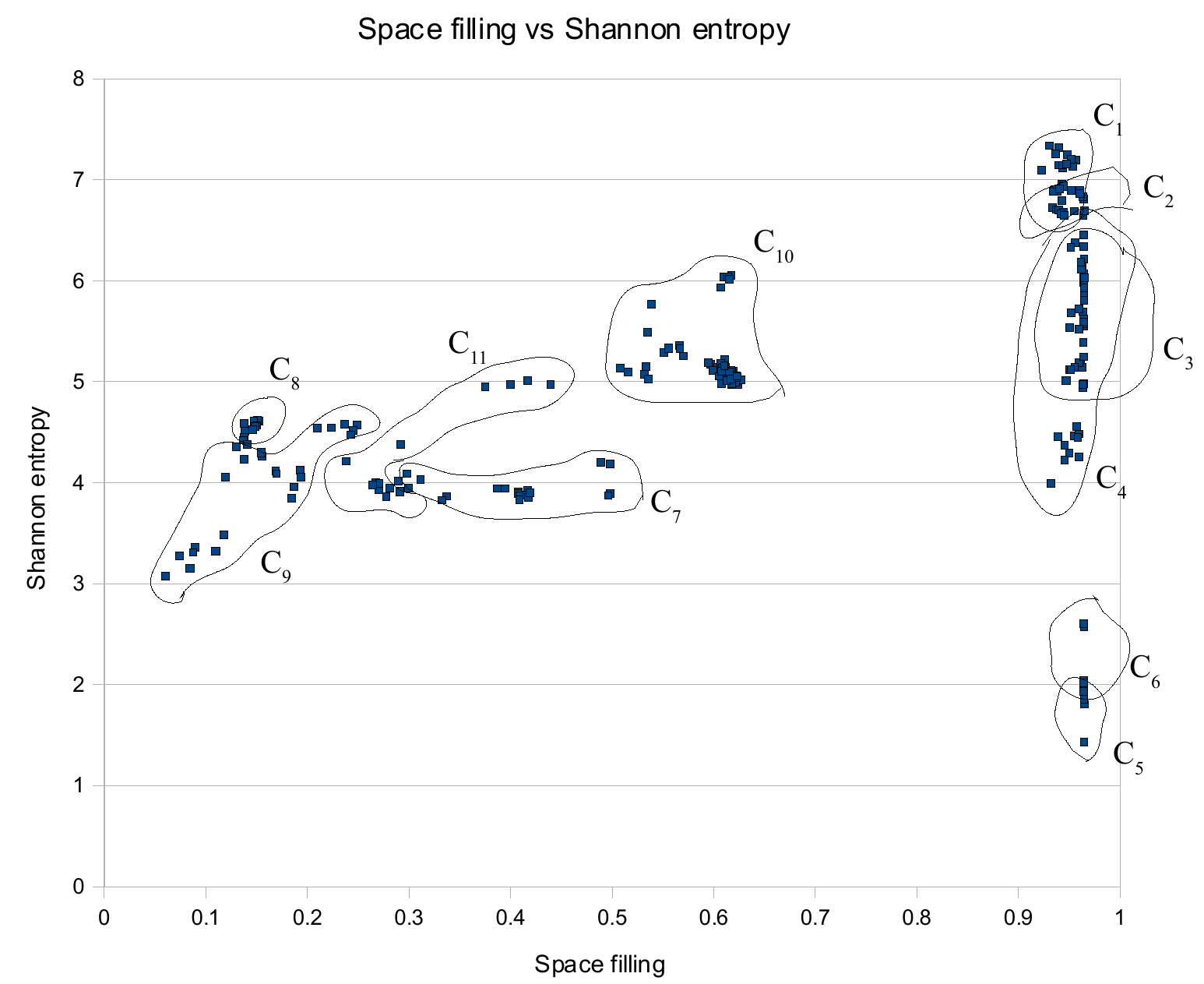}
\caption{Space-filling ratio vs Shannon entropy with $D$-stimulation.}
\label{SpaceFillingShannon}
\end{figure}

\begin{finding}
Classes of excitable cellular automata with retained refractoriness obey the following hierarchy of expressivenes: 
$$C_8 \rhd C_9 \rhd C_{11} \rhd C_{7} \rhd C_{10} \rhd C_1 \rhd C_3 \rhd C_{4} \rhd C_{2} 
\rhd C_{6} \rhd  C_5 .$$
\end{finding} 

Distribution of functions in a space filing versus Shannon entropy space is shown in Fig.~\ref{SpaceFillingShannon}.
Despite having below average diversities classes $C_8$ and $C_9$ occupy top level of the expressivness hierarchy. This is because  their space filling demands are incredibly modest. Apart of a random refractory domain left after $D$-perturbation, we observe only propagating excitation wave-front in $C_8$ and trace leaving localizations in $C_9$. They are followed by $C_{11}$. Automata from $C_{11}$ exhibit varieties of localised and spatially extended excitations.

\begin{figure}[!tbp]
\centering
\subfigure[$\M(1228) \in C_1$]{\includegraphics[scale=0.9]{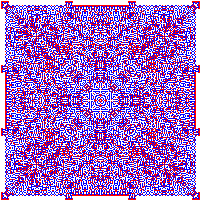}}
\subfigure[$\M(1222) \in C_2$]{\includegraphics[scale=0.9]{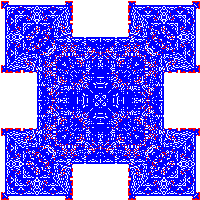}} \\
\subfigure[$\M(2413) \in C_{10}$]{\includegraphics[scale=0.9]{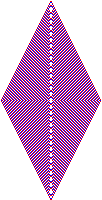}}
\subfigure[$\M(2426) \in C_{10}$]{\includegraphics[scale=0.9]{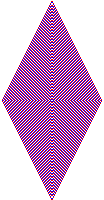}}
\caption{Typical configurations, developed after point-wise excitation, generated by functions 
(a)~$C_1$, (b)~$C_2$, (cd)~$C_{10}$. Names of  functions are indicated in the sub-captions. Snapshots of configurations are taken 100 time steps after point-wise perturbation. Excited cells are red (light grey), refractory cells are blue (dark grey) 
and resting cells are white.}
\label{SingleExcitation_abcd}
\end{figure}

\begin{figure}[!tbp]
\centering
\subfigure[$\M(2211) \in C_9$]{\includegraphics[scale=0.9]{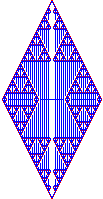}}
\subfigure[$\M(2218) \in C_9$]{\includegraphics[scale=0.9]{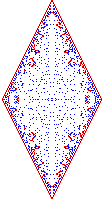}}
\subfigure[$\M(2222) \in C_9$]{\includegraphics[scale=0.9]{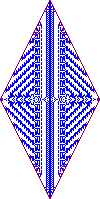}}
\subfigure[$\M(2223) \in C_9$]{\includegraphics[scale=0.9]{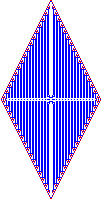}}
\subfigure[$\M(2312) \in C_{11}$]{\includegraphics[scale=0.9]{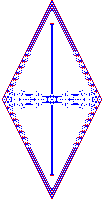}}
\subfigure[$\M(2322) \in C_{11}$]{\includegraphics[scale=0.9]{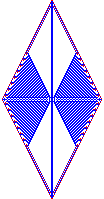}}
\caption{Typical configurations, developed after point-wise excitation, generated by functions 
 (a--d)~$C_9$,
 and (ef)~$C_{11}$. Exact functions are indicated in the sub-captions. Snapshots of configurations are taken 100 time steps after point-wise perturbation. Excited cells are red (light grey), refractory cells are blue (dark grey) 
and resting cells are white.}
\label{SingleExcitation_efghij}
\end{figure}

\begin{figure}[!tbp]
\centering
\begin{footnotesize}
\begin{tabular}{c|cccc}
Class & Shannon entropy &  Simpson's index &  Space-filling ratio & Expressiveness \\ \hline
$C_1$ 	 &6.89	 &1.0 	          &0.90  &    7.65 \\
$C_2$ 	 &5.52 	 &0.98 	 &0.84  &     6.6 \\
$C_3$ 	 &0.31 	& 0.10 	 &1.0    &     0.31 \\
$C_4$ 	 &2.86 	 &0.94 	 &0.08  &     34.32 \\
$C_5$ 	 &0.31 	 &0.1 	          &1.0    &    0.31 \\
$C_6$ 	 &0.31 	& 0.1 	          & 1.0   &     0.31\\
$C_7$ 	 &2.93 	& 0.94 	& 0.09  &     32.58 \\
$C_8$ 	 &2.71 	 &0.93 	& 0.07  &     38.79\\
$C_9$ 	 &4.55 	 &0.96 	& 0.22  &     20.59\\
$C_{10}$ 	 &3.16 	& 0.94 	& 0.27  &     11.71\\
$C_{11}$ 	 &3.80 	 &0.95 	& 0.21  &     17.91\\
\end{tabular}
\end{footnotesize}
\caption{Average values of Shannon entropy, Simpson's index, space-filling ratio and expressiveness calculated on configurations
generated by a point-wise excitation.}
\label{averagesSingleExctiation}
\end{figure}

\begin{finding}
Classes of excitable cellular automata with retained refractoriness obey the following hierarchy of generative diversity: 
$$C_1 \rhd C_2 \rhd C_9 \rhd C_{11} \rhd C_{10} \rhd C_7 \rhd C_4 \rhd C_8 \rhd \{ C_3, C_5, C_6 \} .$$
\end{finding}

Average values of Shannon entropy of configurations generated by point-wise excitation are shown 
in Fig.~\ref{averagesSingleExctiation}. Classes $C_1$ and $C_2$ are at the top of generative diversity hierarchy. Typical configurations developed after a point-wise excitation include almost-rectangular patterns (Fig.~\ref{SingleExcitation_abcd}a) and rectangular patterns with 2/3 diamond corners (Fig.~\ref{SingleExcitation_abcd}b).  These patterns are symmetric yet sophisticated in terms of visual complexity patterns of excited and refractory states.

In the descending order of generative diversity, classes $C_1$ and $C_{2}$ are followed by 
classes $C_9$ and $C_{11}$. Patterns generated in the result of point-wise stimulation include
\begin{itemize}
\item Sierpinski carpets expanding north and south from longest sides of two-cell point-wise excitation 
(Fig.~\ref{SingleExcitation_efghij}a); 
\item rhomboid excitation wave fronts, followed by travelling localizations and leaving 
behind a domain of disordered sparsely distributed refractory states (Fig.~\ref{SingleExcitation_efghij}b); 
\item rhomboid wave-fronts forming ordered strips of refractory states (Fig.~\ref{SingleExcitation_efghij}cd); 
\item and combinations of localised and spatially-extended refractory-state domains (Fig.~\ref{SingleExcitation_efghij}ef).
\end{itemize}

Class $C_{10}$ occupies a mid-range of generative diversity. A point-wise excitation generates rhomboid patterns of  wave packets, each excitation front is followed by a refractory tail (Fig.~\ref{SingleExcitation_abcd}cd). Some functions of the class also produce a narrow band of resting states, surrounded by excitation domain 
(Fig.~\ref{SingleExcitation_abcd}c). Classes $C_3$, $C_5$, $C_4$, $C_6$ $C_8$ have lowest generative diversity. A point-wise excitation leads to formation of a circular excitation wave, which leaves a trail of refractory states.

\begin{figure}[!tbp]
\centering
\includegraphics[width=0.85\textwidth]{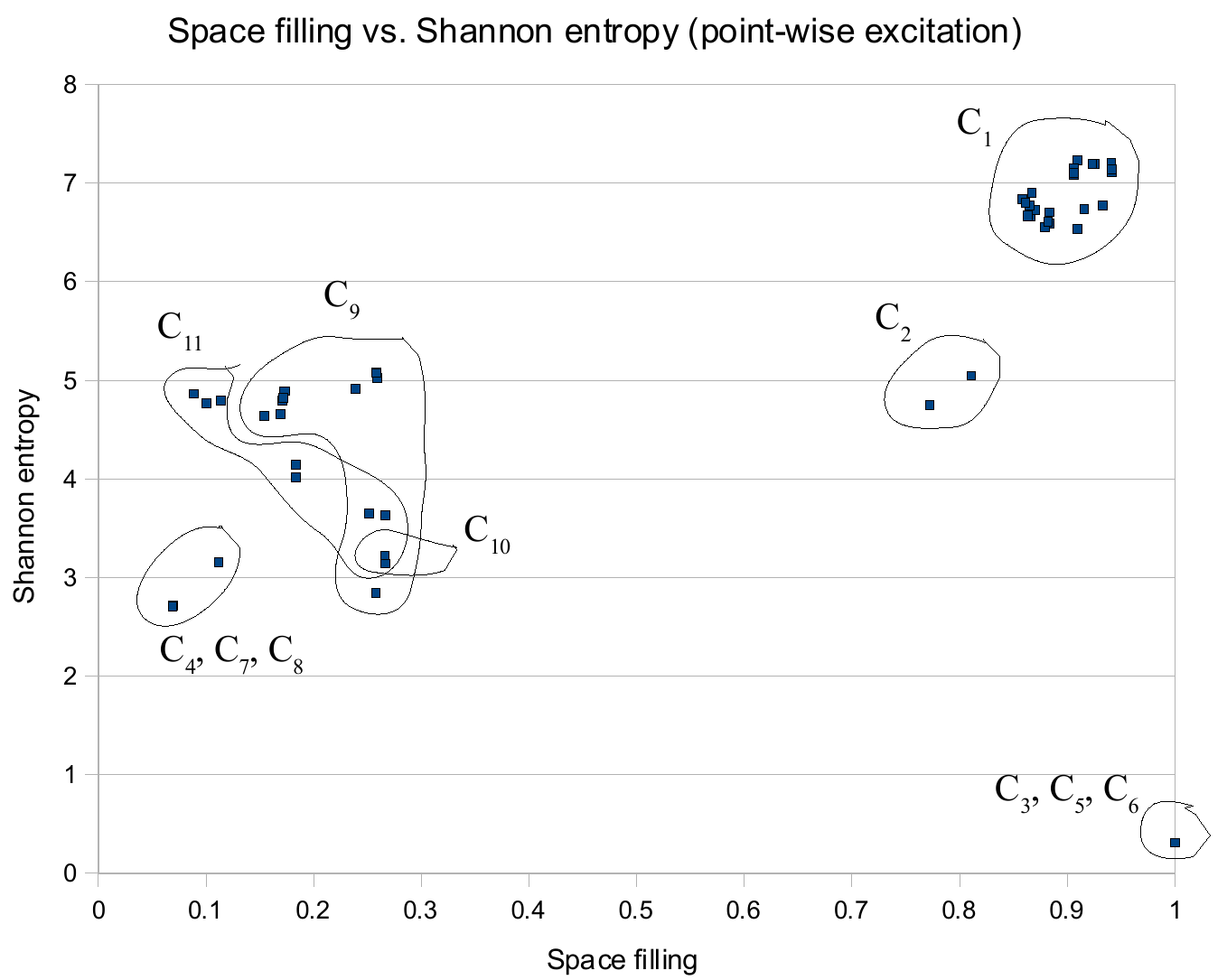} 
\caption{Space-filling ratio vs Shannon entropy. Point-wise excitation.}
\label{SpaceFillingShannonSingleExcitation}
\end{figure}

\begin{finding}
Classes of excitable cellular automata with retained refractoriness obey the following hierarchy of generative expressiveness: 
$$C_8 \rhd C_4 \rhd C_7 \rhd C_9 \rhd C_{11} \rhd C_{10} \rhd C_1 \rhd C_2 \rhd \{ C_3, C_5, C_6 \}. $$
\end{finding} 

See detailed structure of the hierarchy in  Fig.~\ref{SpaceFillingShannonSingleExcitation}.

\section{Travelling localisations}
\label{localisations}

\begin{figure}[!tbp]
\centering
\subfigure[]{\includegraphics[scale=4]{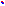}} 
\subfigure[]{\includegraphics[scale=4]{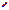}} 
\subfigure[]{\includegraphics[scale=4]{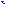}} 
\subfigure[]{\includegraphics[scale=4]{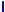}} 
\subfigure[]{\includegraphics[scale=4]{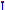}} 
\subfigure[]{\includegraphics[scale=4]{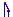}} 
\subfigure[]{\includegraphics[scale=4]{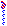}} 
\subfigure[]{\includegraphics[scale=4]{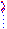}} 
\caption{Examples of travelling localizations found in configurations of automata with retained refractoriness.
(a--e)~Gliders, (f--h)~Trains. Excited cells are red (light grey), refractory cells are blue (dark grey) 
and resting cells are white.}
\label{localisations}
\end{figure}

Travelling  localised excitations is a unique phenomenon of sub-excitable media, widely used in  theoretical and experimental laboratory implementations of novel computing architectures, see 
e.g.~\cite{adamatzky_book_2005}.  Most common localizations found in experiments with excitable automata with retained refractoriness are gliders, trains, glider guns, and spaceships. These structures are well-known in  Conway's Game-of-Life cellular automata (and we are using terminology of the Conway's Game of Life) however they display a range of  uncommon properties when inhabiting discrete excitable medium.  

Examples of gliders (localizations which do not leave traces) and trains (localizations which leave traces made of stationary refractory patterns) are shown in Fig.~\ref{localisations}; all localizations shown there travel north. 
A glider shown in Fig.~\ref{localisations}a consists of three excited and   states. The small gliders can be joined together to make propagating wave-fragments: in Fig.~\ref{localisations}b we see a glider made of three excited and three refractory states joined with a localisation of two excited and two refractory states. The gliders  move along columns and rows of a cellular array.  Similarly to our previous models of sub-excitable discrete media, e.g.~\cite{adamatzky_book_2005}, we found only one excited localisation which travels along diagonals of the cellular array. The diagonal glide consists of three excited and three refractory states, an exemplar snapshot is shown in Fig.~\ref{localisations}c. 

Gliders which leave traces of stationary  patterns behind are called trains. The trains are very common in 
classes $C_9$, $C_{10}$ and $C_{11}$. A minimal train consists of a pair of excited states leaving a column/row of 
refractory pairs behind (Fig.~\ref{localisations}d). In terms of trace minimality, the train shown in 
Fig.~\ref{localisations}e is minimal localisation of three excited states which leaves a trace of refractory states one-cell wide. Trains can be linked with gliders and other trains, an example is shown in (Fig.~\ref{localisations}f). 
Trains leaving localised traces of refractory states are shown in  (Fig.~\ref{localisations}gh).

\begin{figure}[!tbp]
\centering
\subfigure[]{\includegraphics[scale=2.2]{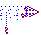}}
\subfigure[]{\includegraphics[scale=2.2]{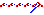}}
\subfigure[]{\includegraphics[scale=2.2]{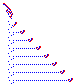}}
\subfigure[]{\includegraphics[scale=2.2]{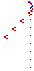}}
\subfigure[]{\includegraphics[scale=2.2]{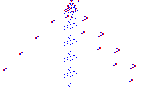}}
\caption{Examples of mobile glider guns found in development of automata 
with retained refractoriness. Excited cells are red (light grey), refractory cells are blue (dark grey) 
and resting cells are white.}
\label{mobileguns}
\end{figure}

Travelling localizations which generate streams of other localizations, e.g. gliders, are called 
mobile  guns (Fig.~\ref{mobileguns}). A mobile gun can emit localisation in any directions, 
subject to the lattice's discreteness, apart of the direction of the gun's motion.  Here we illustrate just few examples:
gun travels east and emits trains travelling south (Fig.~\ref{mobileguns}a),
gun travels east and emits gliders travelling west (Fig.~\ref{mobileguns}b),
gun travels north and emits trains travelling east (Fig.~\ref{mobileguns}c),
gun travels north and emits gliders travelling west (Fig.~\ref{mobileguns}d).  
A complex gun is shown in (Fig.~\ref{mobileguns}e): the gun travels north and emits three streams of gliders, one stream travels west and two streams travel east. 

\begin{figure}[!tbp]
\centering
\includegraphics[scale=2]{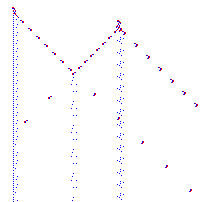}
\caption{Example of interactions between travelling excitations in automata 
with retained refractoriness. Excited cells are red (light grey), refractory cells are blue (dark grey) 
and resting cells are white.}
\label{exampleofinteractions}
\end{figure}

Interactions between gliders is pretty rich, gliders and trains can reflect or annihilate in the result of the collision, also 
refractory stationary patterns can be formed in the results of such collisions. An example is shown in 
Fig.~\ref{exampleofinteractions}.  Two mobile guns travel north. They both leave a trace of refractory states, as trains do.  Westward gun emits a single stream of gliders travelling east. Eastward gun emits two streams of gliders. 
Eastward stream of westward gun collides with westward stream of eastward gun.  Refractory patters are formed in the result of collision, and glider streams are pruned: only every fifth glider of westward stream survives collision and only every fourth glider of eastward stream survives the collision (Fig.~\ref{exampleofinteractions}).

\begin{figure}[!tbp]
\centering
\includegraphics[scale=1.2]{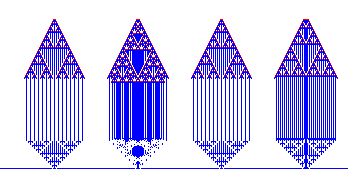}
\caption{'Flying' Sierpinski carpets, generated by automata of class $C_9$. Carpets travel north. 
Patterns (from the left to the right) are generated by lines of two, three, four and five excited cells. 
Launching pad is constructed by a glider, which travels east and leaves a refractory trail.  
The configurations are generated by  function $\M(2311)$. Excited cells are red (light grey), refractory cells are blue (dark grey).}
\label{propercarpets1}
\end{figure}

Functions $\M(2a1b)$, where $a=3, \ldots, 8$ and $b=1,2$, generate spaceships ($b=1$) and Sierpinski flying carpets ($b=2$).  Sierpinski carpets and spaceships observed in many types of cellular automata including Conway's Game of Life~\cite{Wisialowski, LifeWikiSpaceships}. However, we believe, travelling Sierpinski carpets have never been observed before in cellular automata. Examples of Sierpinski carpets generated by function $\M(2311)$ from seeds of different sizes are shown in Fig.~\ref{propercarpets1}. The carpets always leave a trail of refractory states. Therefore they can be considered as analogies of puffer trains~\cite{puffertrain} in the Game of Life cellular automata.

\begin{figure}[!tbp]
\centering
\subfigure[]{\includegraphics[scale=0.9]{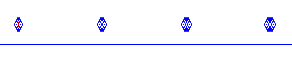}}
\subfigure[]{\includegraphics[scale=0.9]{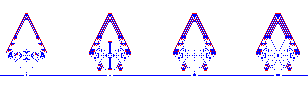}}
\subfigure[]{\includegraphics[scale=0.9]{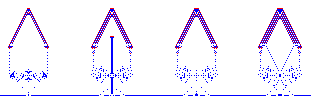}}
\subfigure[]{\includegraphics[scale=0.9]{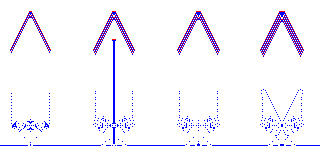}}
\caption{Formation and propagation of spaceships in automata of class $C_{11}$.
 Patterns (from the left to the right) are generated by lines of two, three, four and five 
 excited cells. Launching pad is constructed by glider travelling east and leaving refractory 
 trail. (a)~Generation stage. (bc)~Take-off stage. (d)~Flying. The configurations are generated by 
function $\M(2312)$. Excited cells are red (light grey), refractory cells are blue (dark grey).}
\label{Spaceships1}
\end{figure}

Spaceships  do not leave refractory debris behind.  Several stages of spaceships development are 
illustrated in Fig.~\ref{Spaceships1}. A strip of excited states, e.g. '+', "++', "+++' etc., leads to formation 
of self-similar growing rhomb (Fig.~\ref{Spaceships1}a and Fig.~\ref{SingleExcitation_efghij}ae). 
When the south part of the rhomb gets in touch with a refractory strip it becomes 'infected' with refractory states, 
or frozen, thus only north part of the rhomb continues travelling (Fig.~\ref{Spaceships1}bc). At some stage 
a spaceship becomes disconnected from refractory debris it left and gets 'airborne' (Fig.~\ref{Spaceships1}d). 

Both Sierpinski flying carpets and spaceships observed are rather close relatives of 'grayships'~\cite{LifeWikiSpaceships},
because they are comprised of still patterns which can be extended to arbitrary size. The size of each carpet or 
spaceship is determined by the distance of its seed from the launchpad (a chain of refractory states, which prevents the carpet/spaceship from expanding in one direction).

\begin{figure}[!tbp]
\centering
\subfigure[]{\includegraphics[scale=0.45]{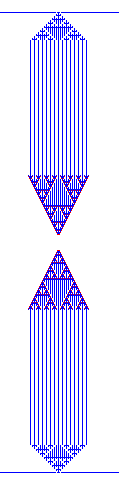}}
\subfigure[]{\includegraphics[scale=0.45]{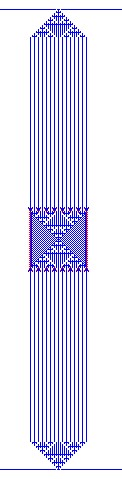}}
\subfigure[]{\includegraphics[scale=0.45]{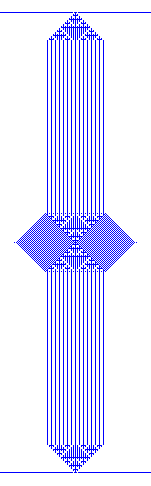}}
\subfigure[]{\includegraphics[scale=0.45]{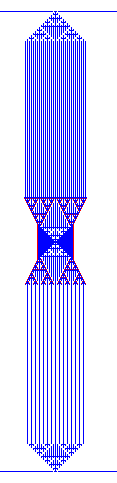}}
\subfigure[]{\includegraphics[scale=0.45]{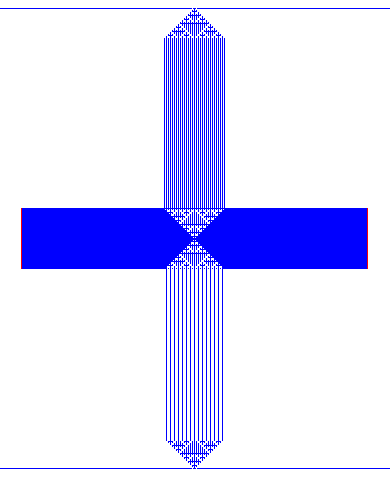}}
\caption{Interaction of flying Sierpinski carpets in automata of class $C_{11}$.
(a)~Carpets approach each other. 
(bc)~Even distance: (b)~full body impact and (c)~formation of debris/annihilation.
(de)~Odd distance: (d)~full body impact and (e)~formation of new travelling localizations.
The configurations are generated by function $\M(2811)$; seed's size is two cells; distance of a seed 
from southern refractory strip (launchpad) is 50 cells. Excited cells are red (light grey), refractory cells are blue (dark grey).}
\label{CarpetCollision}
\end{figure}

Collisions between the flying carpets and spaceships develop along 'classical' scenarios of collisions between travelling localizations: annihilation and reflection, possibly with emergence of new localizations (Fig.~\ref{CarpetCollision}).

\section{Discussion}
\label{discussion}

Aiming to  design a cellular automaton model of a discrete memristive medium we developed a model of 
excitable cellular automaton with retained refractoriness. The model is ideologically similar to our previous model 
of cellular automata with retained excitation~\cite{adamatzky_2007} but the focus shifted to retained refractoriness. 
An excited cell takes refractory state unconditionally, i.e. independently of the number of its excited neighbours, however the refractory cell can remain in its refractory state depending on the number of its excited neighbours. 

A propagation of excitation in a medium can be considered as an analog of electrical current. A refractory state is an automaton analog of high-resistance state of a memristive element, and a switching form the refractory state to resting state, controlled by a density of local excitation, is an analog of switching the memristive element to its state of low-resistance. The automata studied are semi-memristive because switching from low-resistance (resting) state to high-resistance (refractory) state is not controlled by electrical current (density of local excitation). 

Classical automaton models of excitable media are usually based on the principle of threshold excitation: a 
resting cell excites if the number of excited neighbours exceeds certain threshold. This threshold-view on excitation dynamics was the main reason why mobile localizations have not been discovered for so long time. In~\cite{adamatzky_1998} we introduced interval-based excitation: resting cell takes excited state if the number of its excited neighbours  belongs to some specified interval. The approach proved to be productive --- a range of non-classical computing devices~\cite{adamatzky_book_2005}, based on the rich phenomenology of excitation in interval-excitation-based models, was designed. Therefore we employed interval-based excitation and also interval-based recovery, i.e. transition from refractory to resting state, to imitate semi-memristive automata.

Based on functions' behavioural grouping, visual categorisation of configurations generated in a response to random perturbation, ordering functions on their morphological diversity and expressiveness we selected eleven classes of functions: $C_1, \ldots, C_{11}$. 

The functions of classes $C_1, \ldots, C_6$ show high degree of space-filling ratios, while classes $C_7, \ldots, C_{11}$ mid-range and low space-filling ratios.  Low space-filling classes can be further sub-divided into classes with extinct  excitation, namely $C_7$ and $C_8$, where only a single centripetal wave-front is formed,  and classes that support travelling localizations: $C_9, C_{10}, C_{11}$. The low space-filling classes $C_7, \ldots, C_{11}$ also demonstrate higher ratio of excited states over refractory states in configuration functions of the classes generate.  

Top four classes in the hierarchy of morphological diversity are $C_1$, $C_3$, $C_4$ and $C_{10}$. Automata from $C_1$, $C_3$ and $C_4$ show rather quasi-chaotic response to spatially extended random excitation, and this is the reason for their high morphological diversity. Class $C_{10}$ is more interesting. A spatially extended random perturbation leaves a random configuration of refractory states filled with non-expanding, sometimes breathing, domains 
of  localised excitations and a combination of propagating wave-fronts, wave-fragments and travelling localizations. 

Top four classes in the hierarchy of generative diversity are $C_1$, $C_2$, $C_9$ and $C_11$. Classes $C_1$ and $C_2$ show quasi-chaotic dynamics and even a point-wise excitation of a resting lattice leads to formation of a complex growing pattern. Class $C_9$ got to the top four of generative diversity because a point-wise excitations give birth to expanding Sierpinski carpets, and disordered and highly-ordered arrangements of refractory states.

To refine our classification even further we introduced a measure of expressiveness, which relates morphological diversity to space-filling ratio, and thus reflects an 'economy of diversity' in cellular automaton configurations. As you can see in 
Fig.~\ref{SpaceFillingShannon}, classes $C_9, \ldots C_{11}$ show higher degrees of expressiveness. Thus we can propose that functions which support travelling localizations may be at the mid-range levels of morphological diversity but top levels of expressiveness.

We briefly characterised a range of travelling localizations which emerged in the cellular automata studied. We demonstrated that
many localizations are capable for laying sophisticated patterns of refractory, high-resistive in memristor's terminology, states. Such patterns could be used in further studies to compartmentalise otherwise homogeneous spatially extended memristive media. We found outcomes of collisions between propagating localisation are quite rich and therefore all scenarios of collision-based computing~\cite{adamatzky_CBC} can be implemented in semi-memristive cellular automata. 
Our further studies will be concerned with developing fully memristive cellular automaton model and designing laboratory experimental prototypes of large-scale memristive media.



\newpage 

\section*{Bibliography}


\begin{thebibliography}{99}


\bibitem
{adamatzky_1998}	
Adamatzky A. and Holland O. 
Phenomenology of excitation in 2D cellular automata and swarm systems, 
 Chaos, Solitons \& Fractals  3 (1998) 1233--1265.

\bibitem
{adamatzky_book_2001}
Adamatzky A. 
Computing in Nonlinear Media and Automata Collectives (IoP Publishing, London, 2001).


\bibitem{adamatzky_CBC}
Adamatzky A. (Ed.) 
Collision Based Computing 
Springer, 2003.

\bibitem
{adamatzky_book_2005}
Adamatzky A., De Lacy Costello B., Asai T. 
 Reaction-Diffusion Computers (Elsevier, Amsterdam, New York, 2005).

\bibitem{adamatzky_2007}
Adamatzky A. Phenomenology of retained excitation. 
Int J Bifurcation and Chaos  17 (2007) 3985--4014.   

\bibitem{adamatzky_memristive_excitable}
Adamatzky~A. and Chua~L. 
Memristive excitable cellular automata.
Int. J. Bifurcation Chaos (2011), in press.

\bibitem
{chua:1971}
 Chua~L.~O., Memristor --- the missing circuit element.
 IEEE Trans. Circuit Theory 18 (1971) 507--519.
 
 \bibitem
 {chua:1976}
 Chua~L.~O.  and Kang~S.~M., 
 Memristive devices and systems. 
 Proc. IEEE 64 (1976) 209--223.
 
 \bibitem
 {chua:1980} 
Chua~L.~O. 
Device modeling via non-linear circuit elements.
 IEEE Trans. Circuits Systems 27 (1980) 1014--1044.

\bibitem
{erokhin:2008}
 Erokhin V., Fontana M.T. Electrochemically controlled polymeric device: a memristors (and more) found two years ago. (2008) arXiv:0807.0333v1 [cond-mat.soft]



\bibitem
{halpern:1990}
Halpern~P. and G. Caltagirone.
Behavior of topological cellular automata.
Complex Systems 4 (1990) 623Ð651.

\bibitem
{ilachinsky:halpern:1987}
Ilachinsky~A. and Halpern~P., Structurally dynamic cellular automata,
Complex Systems 1 (1987) 503--527.

\bibitem
{itoh:2009}
Itoh~M. and Chua~L. 
Memristor cellular automata and memristor discrete-time cellular neural networks.
Int. J. Bifurcation and Chaos 19 (2009) 3605--3656. 


\bibitem
{greenberg:1978}
Greenberg~J. M.  and Hastings~S. P.  
Spatial patterns for discrete models of diffusion in excitable media, 
SIAM J. Appl. Math. 34 (1978) 515Ð-523.




\bibitem{puffertrain}
Puffer train. LifeWiki. (2010) 
\url{http://www.conwaylife.com/wiki/index.php?title=Puffer_train}

\bibitem{LifeWikiSpaceships}
Spaceship. LifeWiki. (2010) \url{http://www.conwaylife.com/wiki/index.php?title=Spaceship}

\bibitem
{strukov:2008}
Strukov, D.B., Snider, G. S., Stewart, D. R. and Williams, R. S., The missing memristor found. Nature 453 (2008) 80--83.


\bibitem
{williams:2008}
Williams~R.~S. 
How we found the missing memristor. IEEE Spectrum 
2008-12-18.


\bibitem{Wisialowski}
Wisialowski~B.
ConwayÕs Game of Life: Lines to Fractals
\url{https://webfiles.uci.edu/bwisialo/www/gameoflife2.html}

\bibitem
{yang:2008}
Yang, J.J., Pickett, M. D., Li, X., Ohlberg, D. A. A., Stewart, D. R. and Williams, R.S.  Memristive switching mechanism for metal-oxide-metal nanodevices. Nature Nano, 2008 3(7). 


\end{thebibliography}
\end{document}